\documentclass[twocolumn,twoside,slac_two]{revtex4}
\usepackage{graphicx}
\usepackage{fancyhdr}
\begin{document}

\title{Results from the Cryogenic Dark Matter Search experiment}

%

\author{Oleg Kamaev for the CDMS Collaboration}
\affiliation{School of Physics \& Astronomy, University of Minnesota, Minneapolis, MN 55455, USA}
\begin{abstract}
The Cryogenic Dark Matter Search experiment uses low-temperature solid-state detectors to seek Weakly Inter\-acting Massive Particles (WIMPs) and has the world's best exclusion limit on the spin-independent WIMP-nucleon cross section. The experiment uses ionization and athermal phonon signals from particle interactions to discriminate between candidate (nuclear recoil) and background (electron recoil) events with extremely high efficiency. The detectors' low energy threshold for electron recoil events allows us to perform the search for relic axions and solar axions which can interact in the detector via the axio-electric coupling $g_{a\bar{e}e}$ and the Primakoff coupling $g_{a\gamma\gamma}$ respectively. We describe the experiment and our most recent results from the Soudan 5-tower data runs which include a world-leading upper limit on the WIMP-nucleon spin-independent cross section with a minimum of $4.6\times10^{-44}$ cm$^{2}$ at the 90\% confidence level (CL) for a 60 GeV/c$^{2}$ WIMP, the world-leading experimental upper limit on the axio-electric coupling of $1.4\times10^{-12}$ at the 90\% CL for a 2.5 keV/c$^{2}$ axion, and the upper limit on the axion-photon coupling of $2.4\times10^{-9}$ GeV$^{-1}$ at the 95\% CL.
\end{abstract}

\maketitle

\thispagestyle{fancy}


\section{The CDMS experiment}
\label{sec:cdms_experiment}
The Cryogenic Dark Matter Search (CDMS) experiment is designed to search for WIMPs through their elastic scattering with nuclei. The experiment is located at the Soudan Undeground Lab at 2090 meters of water equivalent (m.w.e.) depth. Nineteen Ge (250~grams each) and 11 Si (100~grams each) detectors are mounted into five towers. Each tower consists of six vertically stacked detectors. Each detector is a high-purity Ge or Si crystal in the shape of a 1~cm thick, 7.6~cm diameter disk and contains four photolithographically patterned phonon sensors on one side and two concentric ionization electrodes on the other. The detectors are operated at $\sim$40 mK to collect the athermal phonons caused by particle interactions in the crystal. A combination of active and passive shielding is used to reject events caused by the cosmogenic muons and to reduce the external environmental radioactivity. A detailed description of the CDMS apparatus is given in~\cite{cdms_PRD2005}.  

At the heart of the experiment are Z(depth)-sensitive Ionization Phonon (ZIP) detectors which measure the ionization and the phonon signal from the interaction of particles with the crystal. External gammas and beta-particles interact with an electron in the crystal and such events are called ``electron recoils,'' whereas neutrons and WIMPs interact with a nucleus (``nuclear recoils''). The main signature of the nuclear recoil events is that they produce $\sim$1/3 fewer charge pairs than the electron recoils. Four independent phonon sensors provide phonon energy and position information for the event. Inner and outer electrodes on the ``ionization side'' veto events from the outer part of the detector and provide inner ionization energy measurement. The independent simultaneous measurements of the phonon and ionization energies of an event allow us to discriminate between nuclear and electron recoils by examining the ratio of ionization to the phonon recoil energy (the ``ionization yield parameter''). The ionization yield provides a primarily electron recoil rejection factor of $>10^{4}$, which is as high as $10^{6}$ when combined with timing information.

Passive and active shielding surround the icebox where the ZIP detectors and the cold hardware are located. Passive shielding consists of layers of lead to reduce external gammas with the inner layer made of ultra-low-activity ancient lead, and polyethylene to moderate neutrons from fission as well as from $(\alpha,n)$ processes due to U/Th decays. The active shielding consists of a muon veto system to reject events from cosmogenic muons or showers. Extensive Monte Carlo simulations of nuclear recoil events that are caused by neutrons due to radioactive processes or cosmogenic muons give an upper limit of $<0.1$ events from each source for the WIMP-search data presented in the next section.  
   
\section{Data analysis and WIMP-search results}
\label{sec:wimp_search}
\begin{figure}[t]
\centering
\includegraphics[width=80mm]{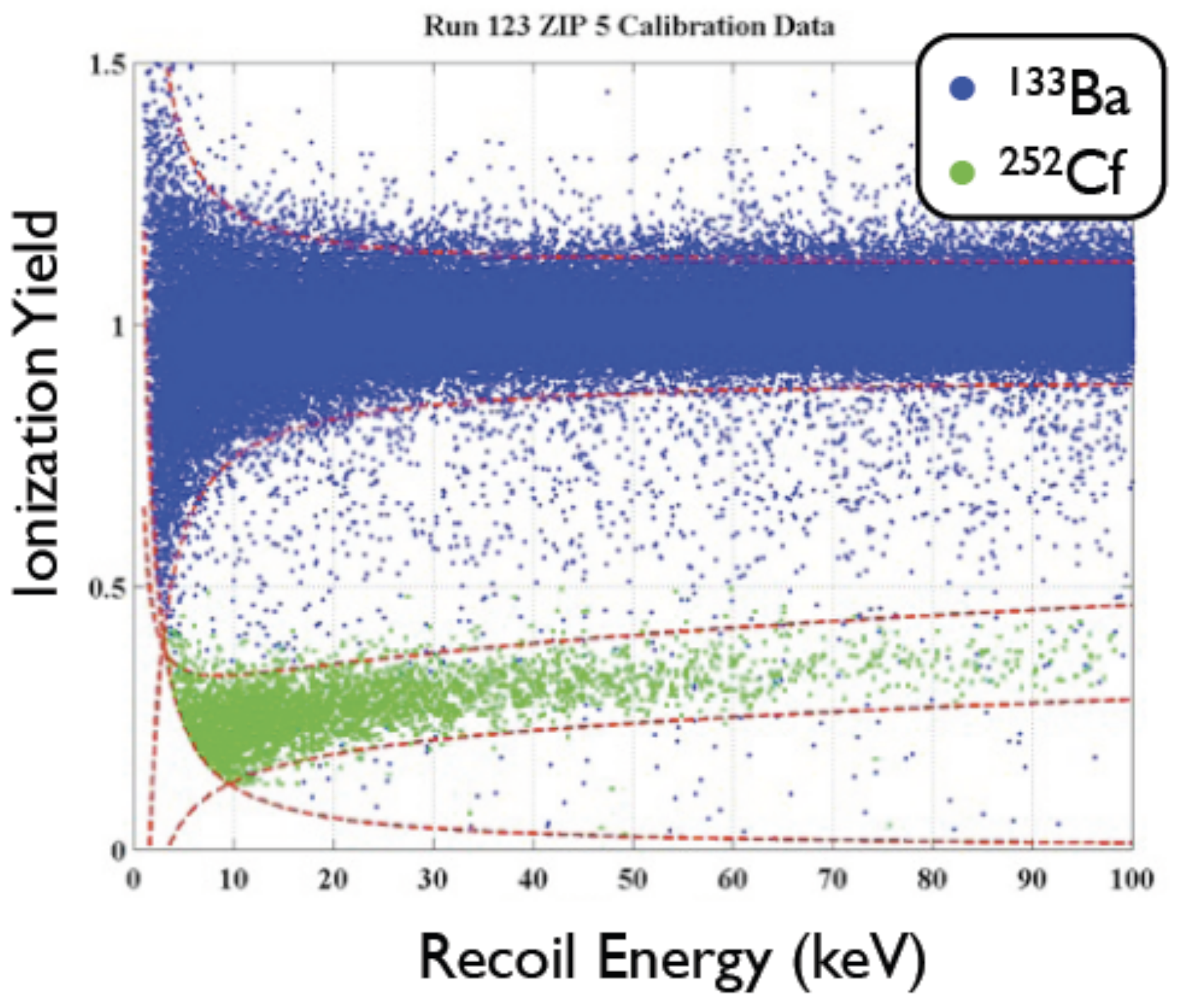}
\caption{Ionization yield vs. the full recoil energy. Electron recoil band (blue dots, upper band) and nuclear recoil band (green dots, lower band) are defined with $^{133}$Ba and $^{252}$Cf  calibration sources respectively.} 
\label{fig:yield}
\end{figure}

Data collected between October 2006 and July 2007, which correspond to cryogenic runs R123--R124, were analyzed and the analysis summary is presented here. For the analysis, 15 good Ge detectors were used in R123 and 7 good Ge detectors in R124 which gave a total raw exposure of 397.8 kg-days. Calibration data with $^{133}$Ba and $^{252}$Cf radioactive sources were taken periodically to characterize the detectors performance. They were also used to define the electron and nuclear recoil bands (Fig.~\ref{fig:yield}) and for WIMP-search efficiency studies. 

\begin{figure}[t]
\centering
\includegraphics[width=80mm]{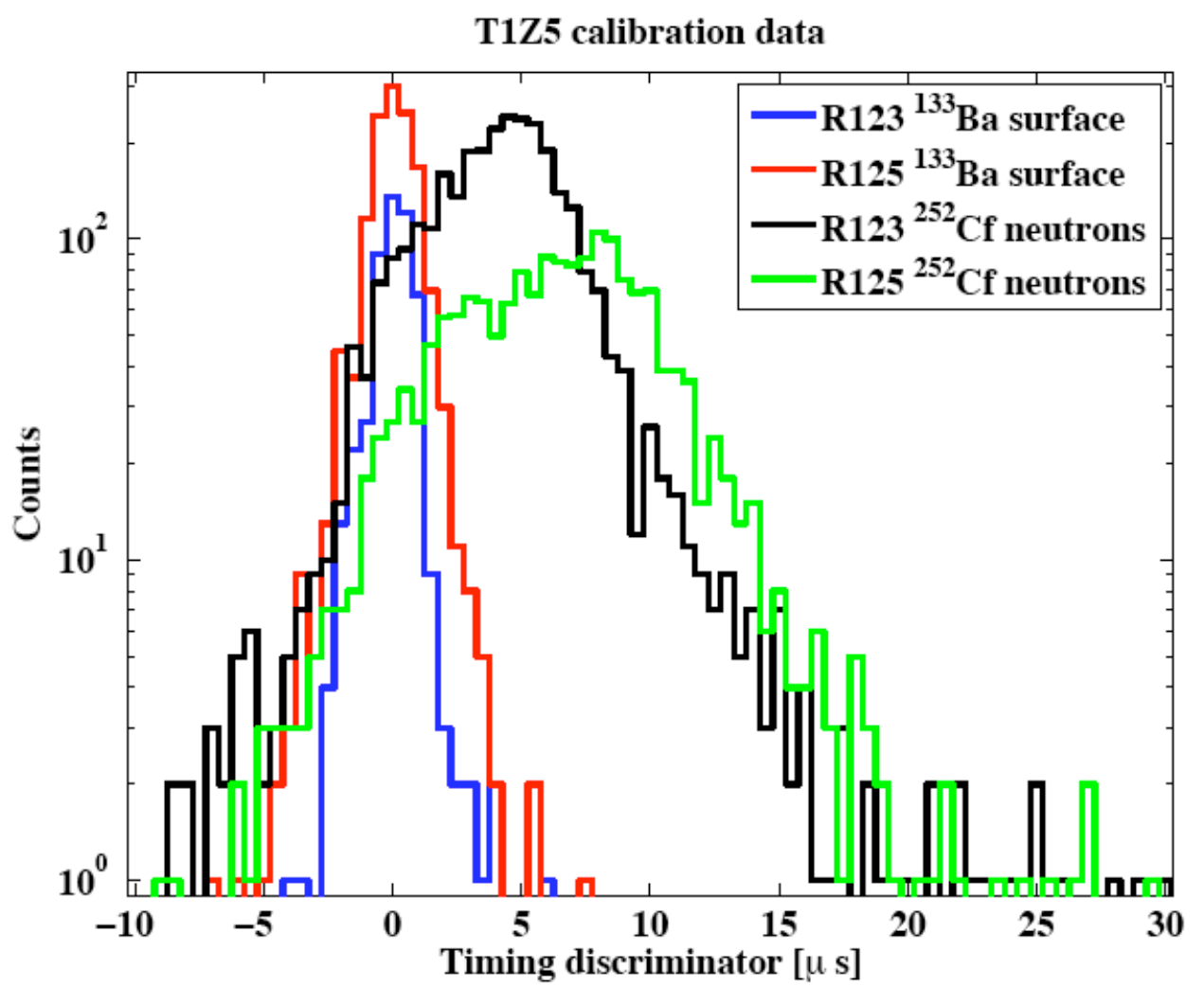}
\caption{Timing parameter distributions for neutrons from $^{252}$Cf and low-yield events from $^{133}$Ba.} 
\label{fig:timing}
\end{figure}

The dominating background for the analysis consists of low-yield electron recoil events, also called ``surface'' events. They originate from interactions that occur in the first few microns of the crystal surfaces, have suppressed ionization signal, and can be misidentified as nuclear recoils. The timing characteristics of the phonon pulse can be used to discriminate surface events. A linear combination of the leading phonon pulse risetime with the delay of the phonon signal with respect to the ionization signal was employed, and its distribution for the calibration data is shown in Fig.~\ref{fig:timing}. Surface events have smaller risetime and delay values. The surface event rejection factor from the timing quantity is $>200$, which improves the overall rejection of electron recoils to $>10^{6}$.

WIMP-search candidate events were required to be inside the 2$\sigma$ nuclear-recoil band in ionization yield, to be veto-anticoincident single-scatters within the detector fiducial volume, and to satisfy data quality, noise rejection, and phonon timing cuts. Detection efficiency was calculated to be $\sim$31\% and gave a total spectrum averaged exposure of 121 kg-days after all cuts for a 60 GeV/c$^{2}$ WIMP. The estimated number of surface events that leak into the signal region, based on the observed number of single- and multiple- scatter events within and surrounding the 2$\sigma$ nuclear-recoil band, is $0.6^{+0.5}_{-0.3}(stat)^{+0.3}_{-0.2}(syst)$ events~\cite{CDMS-09}. 

The WIMP-search blinded signal region was unmasked after all analysis cuts were finalized. No events were observed, and the 90\% CL upper limit on the spin-independent WIMP-nucleon cross section was set with a minimum of $6.6\times10^{-44}$ cm$^{2}$ for a 60 GeV/c$^{2}$ WIMP with the current analyzed dataset ($4.6\times10^{-44}$ cm$^{2}$ when combined with previous CDMS data).  


\subsection{Completing the 5-tower data run}
Data collected between July 2007 and September 2008 correspond to the four cryogenic runs R125--R128 and complete the 5-tower setup data-taking. The raw exposure for these four runs is  $\sim$1.7 times larger than the exposure of R123--R124. 
Improvements made since the previous analysis include a new, faster data processing package written in C++, better timing-based position reconstruction for low-energy events, functional form time-domain fit for phonon pulses, and automation of several data quality cuts. Timing discrimination of surface events looks good compared to the previous analysis and is shown in Fig.~\ref{fig:timing}. 

A significant improvement was achieved for the process of setting the cut to allow a chosen number of surface events to leak into the signal region. An optimization technique was developed to vary the cut on each detector to maximize the total exposure-weighted neutron efficiency while keeping the number of the total leaked events the same. Systematic differences between Ba calibration and WIMP-search data are accurately taken into account. The analysis is in progress and the final results are expected soon. 

\section{Low-energy electron-recoil spectrum analysis}
\label{sec:low_energy}
The DAMA collaboration reported an observation of an excess in the detection rate in the low energy spectrum together with the modulation signal centered at $\sim$3 keV~\cite{DAMA-08}. Interpretation of the result as nuclear recoil interactions requires non-trivial explanations (e.g.~\cite{Pierce-09}) to reconcile with the null result from other experiments~\cite{CDMS-09, XENON-08}. However, DAMA's result may be interpreted as the conversion of dark matter particles into electromagnetic energy in the detector since the DAMA detector does not discriminate between electron and nuclear recoils. In such a case it should be possible to observe the corresponding signal in the CDMS electron recoil band.

The same dataset (R123 and R124) as described in section~\ref{sec:wimp_search} was analyzed with the addition of those Ge detectors that were excluded for the WIMP-search analysis due to their poor surface event discrimination, but are adequate for the electron-recoil spectrum analysis. This increased the total raw exposure to 443.2 kg-days. Events were required to be inside the 2$\sigma$ electron-recoil band in ionization yield. Other selection criteria were similar to the main WIMP-search analysis and required events to be veto-anticoincident single-scatter within the detector fiducial volume satisfying data quality and noise rejection requirements. Detection efficiency varied from 30\% to 70\% depending on the detector. 

\begin{figure}[t]
\centering
\includegraphics[width=80mm]{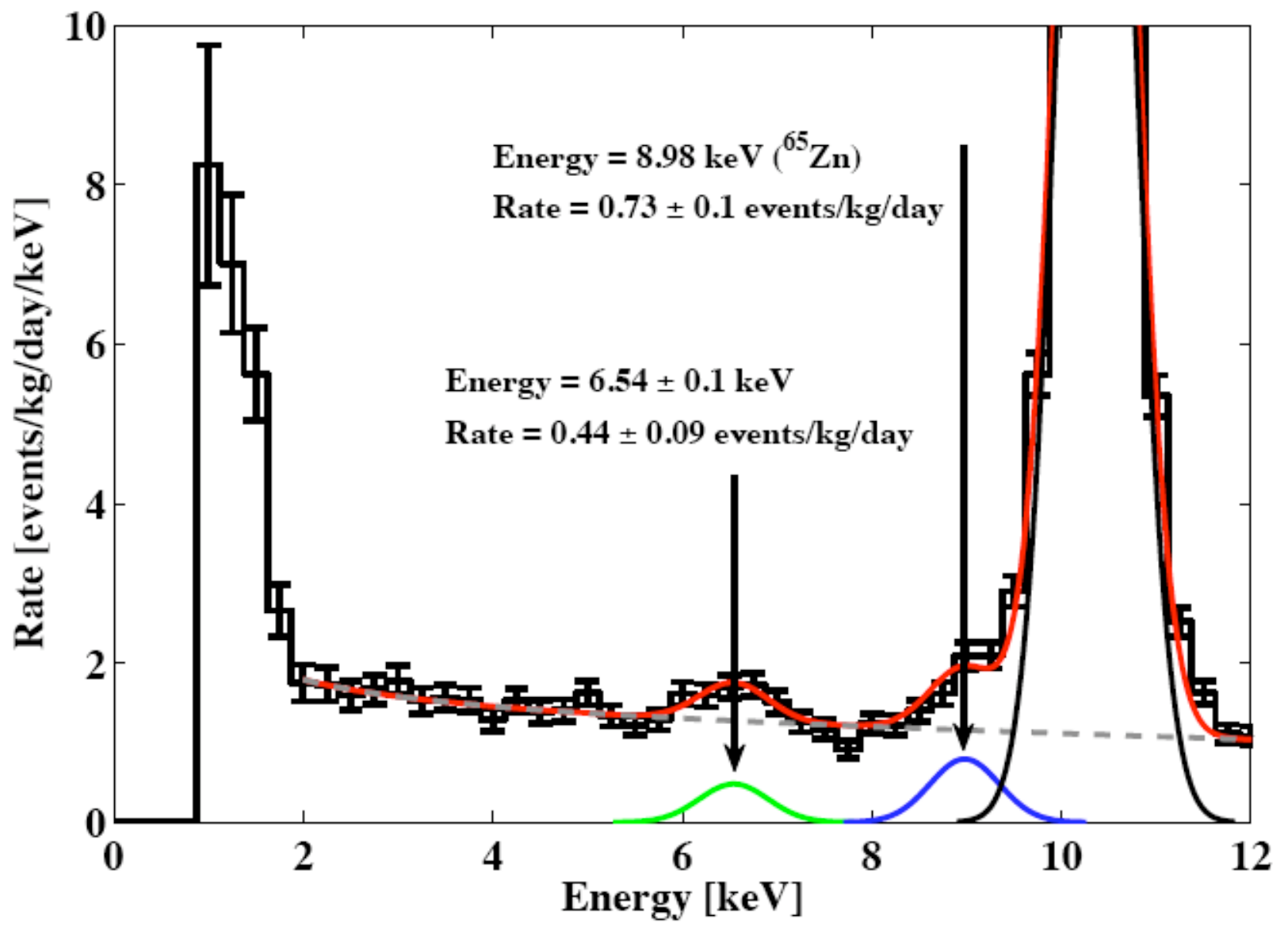}
\caption{Efficiency corrected low-energy spectrum with the fit (red line) as described in the text.} 
\label{fig:low_energy_spec}
\end{figure}

An efficiency-corrected, coadded low-energy spectrum is shown in Fig.~\ref{fig:low_energy_spec}. It is important to note that for this analysis, an electron-equivalent energy range of 2--8.5 keV based on the ionization signal was considered. The spectrum was fit with a sum of a background model and three Gaussian distribution functions that correspond to the known spectral lines at 8.98 keV ($^{65}$Zn decay, remnant from cosmogenic activation), 10.36 keV (X-ray and Auger-electrons from $^{71}$Ge decay), and the line at 6.54 keV which is most likely due to de-excitation of $^{55}$Mn caused by electron-capture decay of $^{55}$Fe formed as the result of cosmogenic activation of germanium.  Note, that 8.98 and 10.36 keV lines are outside of the analysis window. Background rate is $\sim$1.5 counts/kg/day/keV. 

An unbinned log-likelihood fit was used to estimate the excess of event rate above the background. The event rate fit function was a simple sum of the Gaussian distribution function (signal) and the background model together with the Gaussian distribution for the 6.54 keV line multiplied by a weighting factor. The uncertainties in the production rate of $^{55}$Fe and in the time that the detectors spent above the ground, prevent an accurate estimation of the $^{55}$Mn contribution to the spectrum. Thus, the weighting factor is needed to suppress the importance of the 6.54 keV line in the background. By varying it, the most conservative limit on the excess of event rate was taken. The result of the fit indicated that there is no significant excess in the rate above the background.   

\begin{figure}[t]
\centering
\includegraphics[width=80mm]{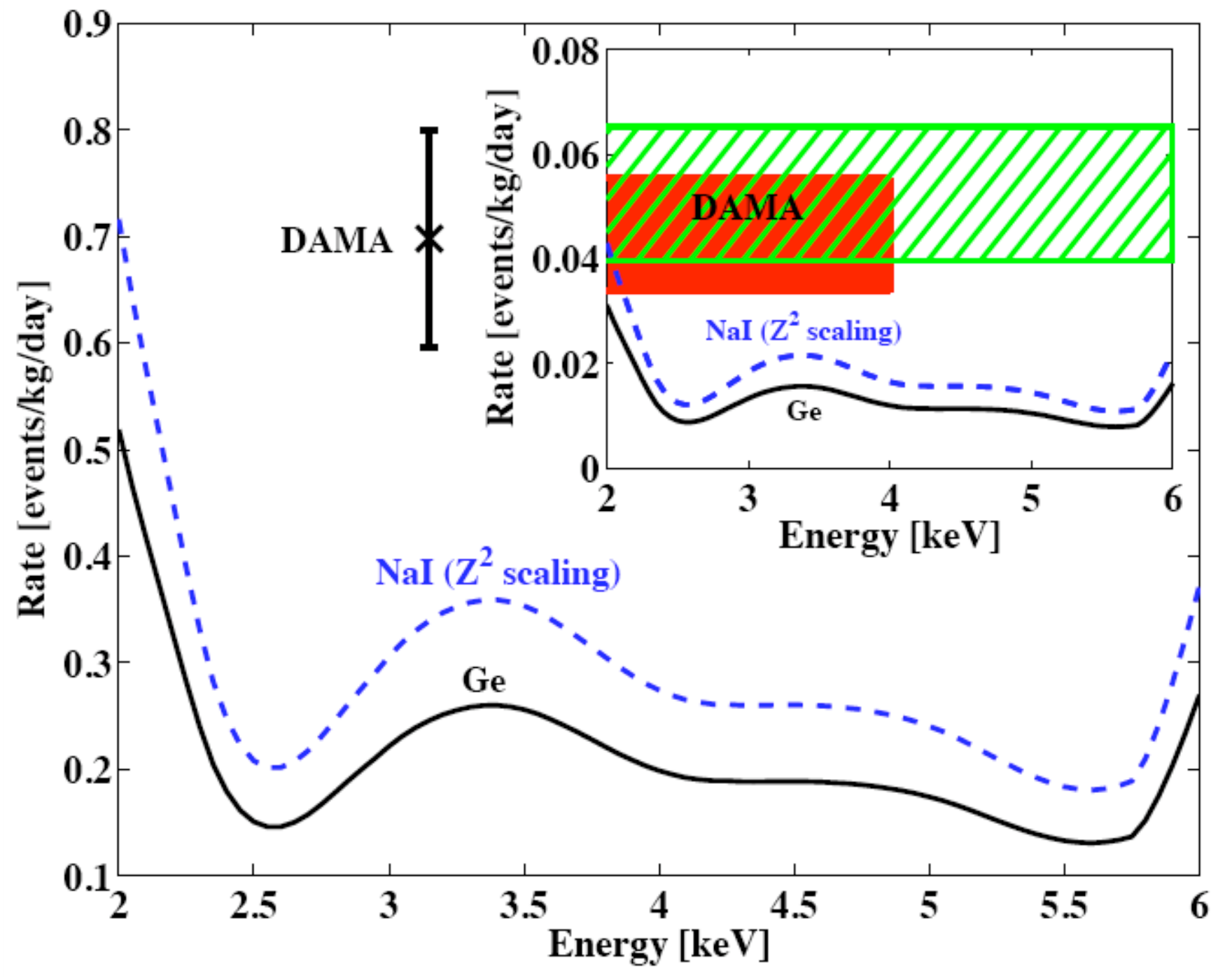}
\caption{Upper limit at the 90\% C.L. on event rate excess above the background together with the DAMA result (data point with error bars). The insert shows the upper limit on the modulation amplitude and compares them with the DAMA modulation signal in 2--4 keV (red, filled) and 2--6 keV (green, hatched) ranges.} 
\label{fig:low_energy_UL}
\end{figure}

An upper limit at the 90\% C.L. on event rate excess above the background is shown in Fig.~\ref{fig:low_energy_UL}~\cite{cdms_low_energy} together with the naive $Z^{2}$ scaling of the limits in Ge to the expected rate in NaI to be able to compare with the DAMA experiment. At 3.15 keV the upper limit curve is 6.8$\sigma$ below the DAMA result. The insert in Fig.~\ref{fig:low_energy_UL} compares the upper limit on the modulation amplitude which is assumed to be 6\% for the standard halo assumption with the 2$\sigma$ region of the annual modulation from DAMA in 2--4 keV and 2--6 keV ranges. Upper limits on the modulation are approximately half the value reported by DAMA. 

\section{Axion search}
The low energy electron-recoil spectrum can also be searched for a signal from the axion-like dark matter particles. These might be relic axions, for which we can set an upper limit on the axio-electric coupling $g_{a\bar{e}e}$, or solar axions, for which we set an upper limit on the Primakoff coupling $g_{a\gamma\gamma}$. Events were selected following the same requirements as for the low-energy electron recoil spectrum analysis described in section~\ref{sec:low_energy}.

\begin{figure}[t]
\centering
\includegraphics[width=80mm]{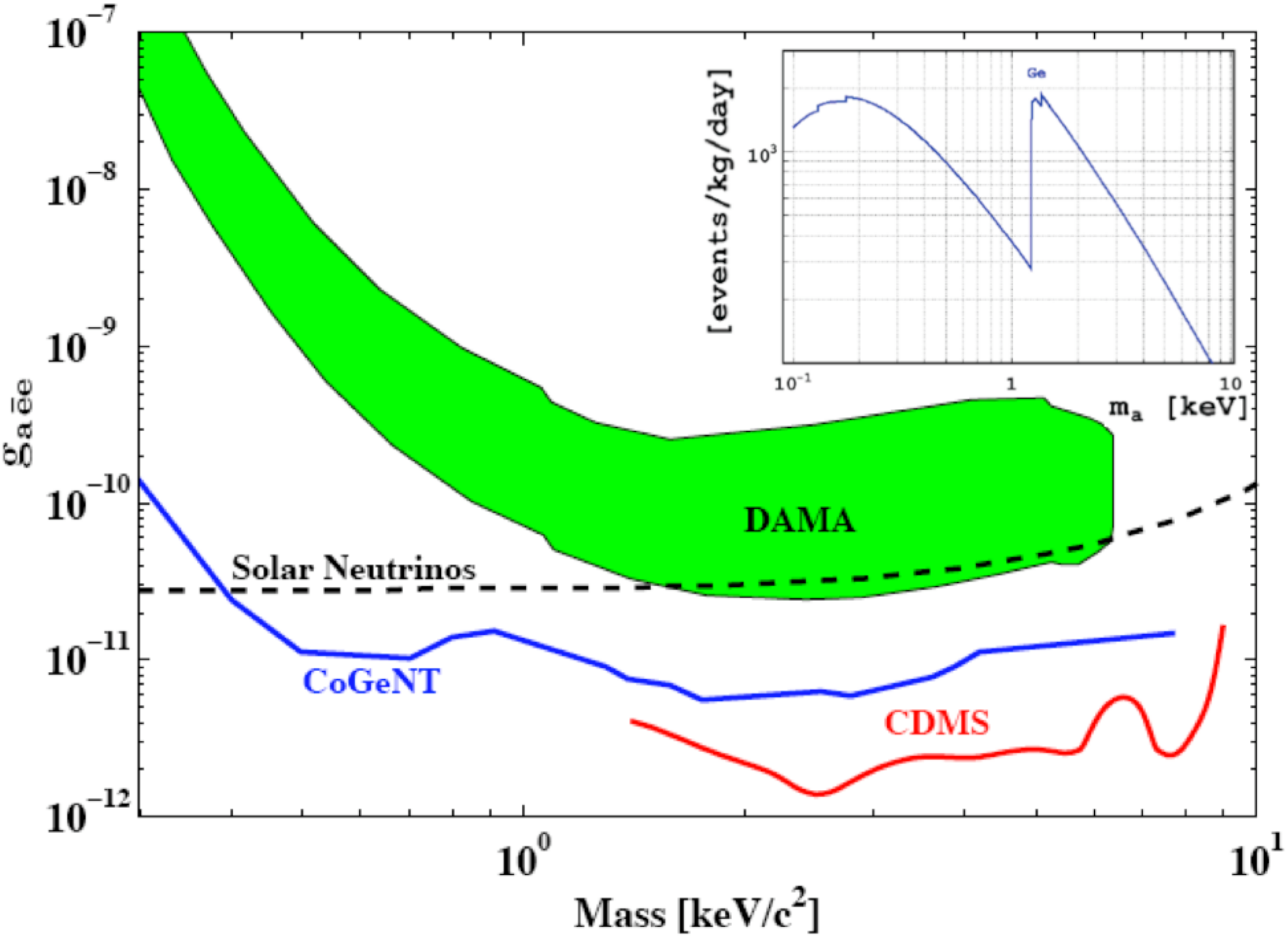}
\caption{The 90\% C.L. upper limit on the axio-electric coupling together with the other experiment results. The insert shows the expected event rate for germanium by the axio-electric coupling for $g_{a\bar{e}e}=10^{-10}$.} 
\label{fig:coupling_aee}
\end{figure}

For the low mass axion ($\sim$keV), pair production is kinematically forbidden. Thus, when interacting with the crystal, the axion is absorbed by a bound electron, which is then ejected from the atom, similar to the photoelectric effect. The interaction rate of the axion-like dark pseudoscalar is proportional to $A^{-1}g^{2}_{a\bar{e}e}\sigma_{p.e.}$, where $A$ is the atomic mass number~\cite{Pospelov:2008jk}. An expected event rate for germanium by the axio-electric coupling for $g_{a\bar{e}e}=10^{-10}$ is shown in the insert of Fig.~\ref{fig:coupling_aee}. 

The expected observable from the interaction of axions with the Ge crystal is the peak at energy $m_{a}$ in the electron-recoil spectrum. The same profile likelihood calculation described in section~\ref{sec:low_energy} was used to set the upper limit on the axio-electric coupling in the absence of a statistically significant excess of event rate above the background. The 90\% C.L. upper limit on the coupling is shown in Fig.~\ref{fig:coupling_aee}~\cite{cdms_axions} together with the allowed region claimed by the DAMA experiment~\cite{Bernabei:2005ca} as a possible galactic axion interpretation of their signal\footnote{For comments on the DAMA allowed region see~\cite{Pospelov:2008jk}.}.  

\begin{figure}[h]
\centering
\includegraphics[width=80mm]{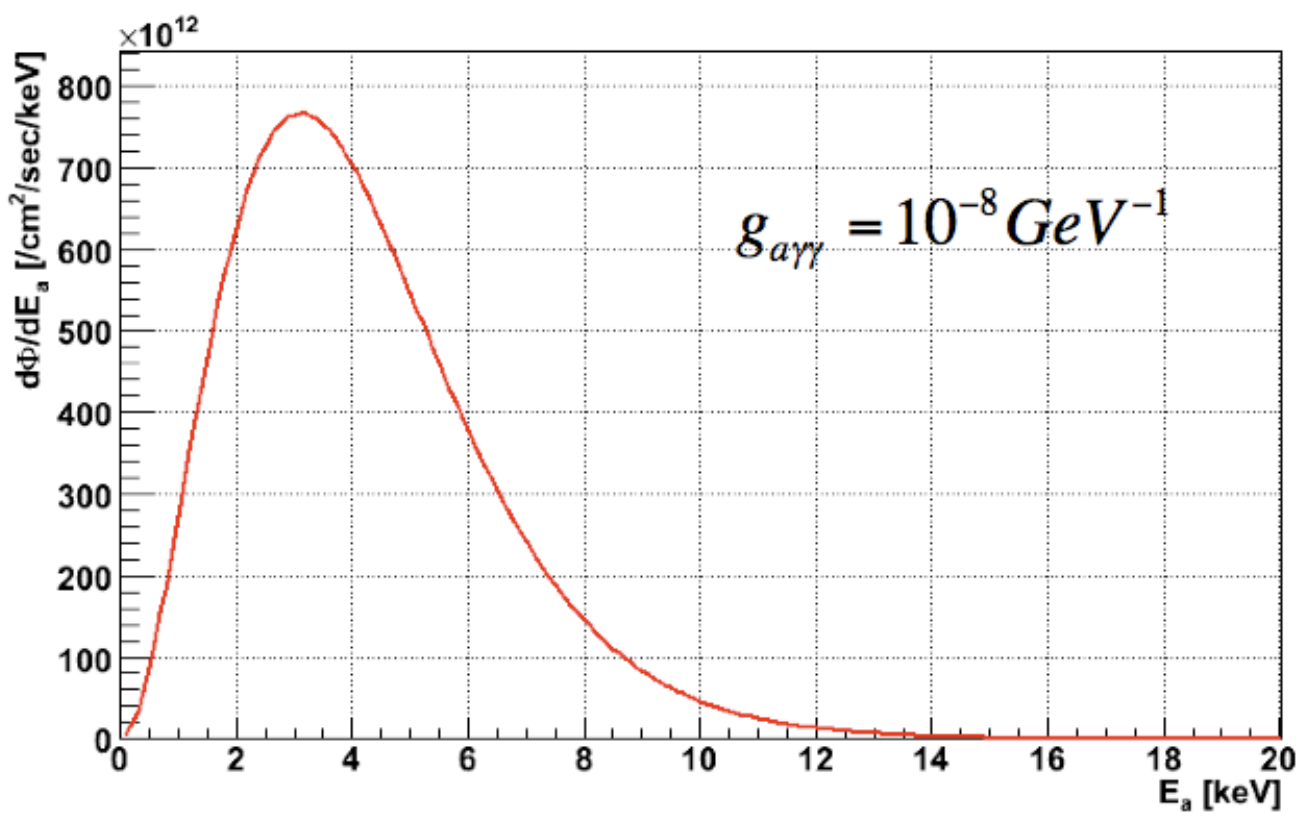}
\caption{Estimated spectrum of the solar axion flux at Earth.} 
\label{fig:solar_flux}
\end{figure}

Estimated spectrum of the solar axion flux at Earth for a given axion-photon coupling is shown in Fig.~\ref{fig:solar_flux}. For axions interacting with a Ge crystal, intense electric field  in the proximity of the nucleus can trigger axion conversion to a photon by the Primakoff effect. Light axions will experience Bragg scattering in a crystal, which implies that the axion energy is inversely proportional to the product of the reciprocal lattice vector and the direction to the Sun. Thus, correlation of the expected rate with the position of the Sun is a signature of the solar axion signal. 
 
\begin{figure}[t]
\centering
\includegraphics[width=80mm]{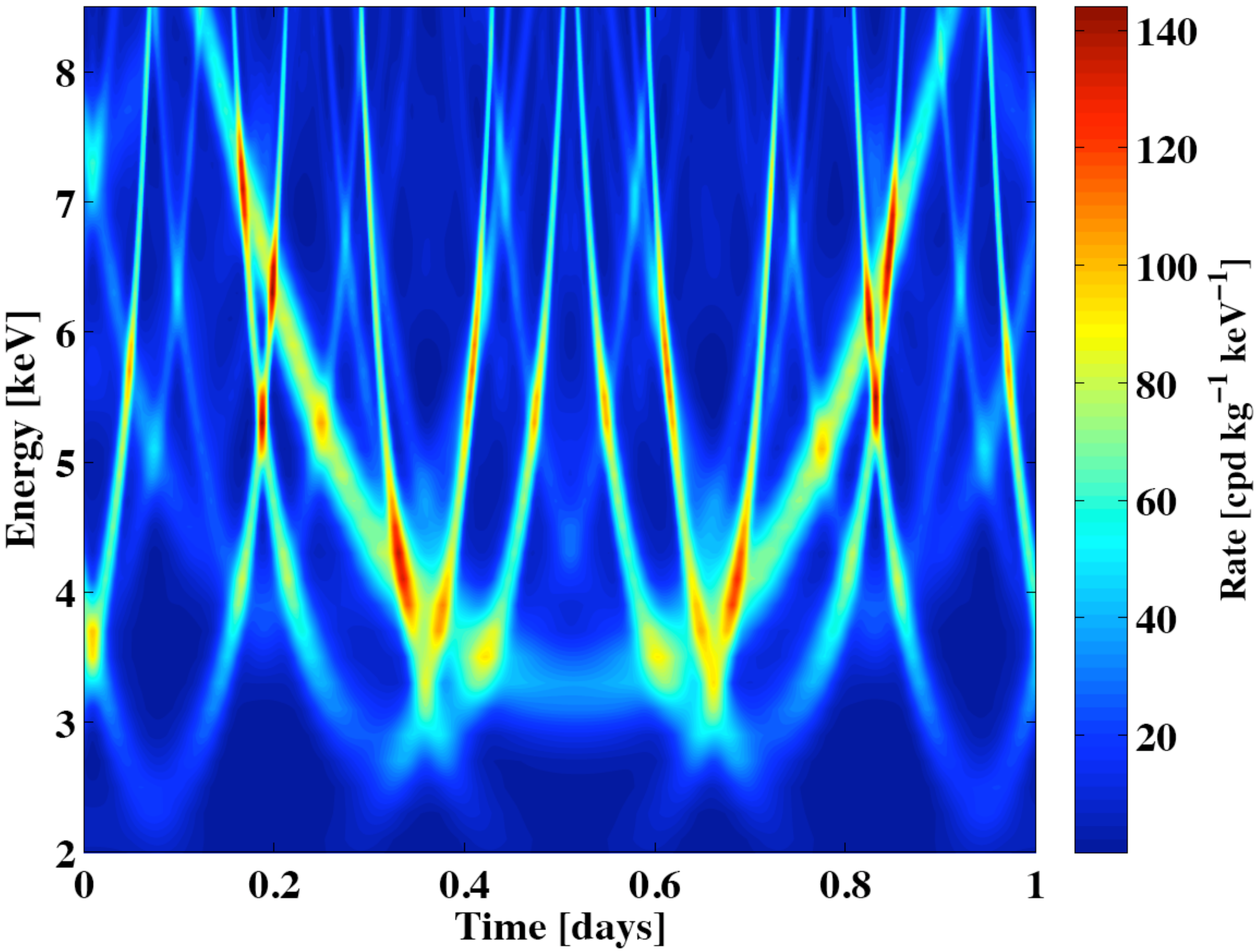}
\caption{Calculated expected solar axion event rate in Ge for $g_{a\gamma\gamma}=10^{-8}$ GeV$^{-1}$.} 
\label{fig:solar_rate}
\end{figure}
\begin{figure}[b]
\centering
\includegraphics[width=80mm]{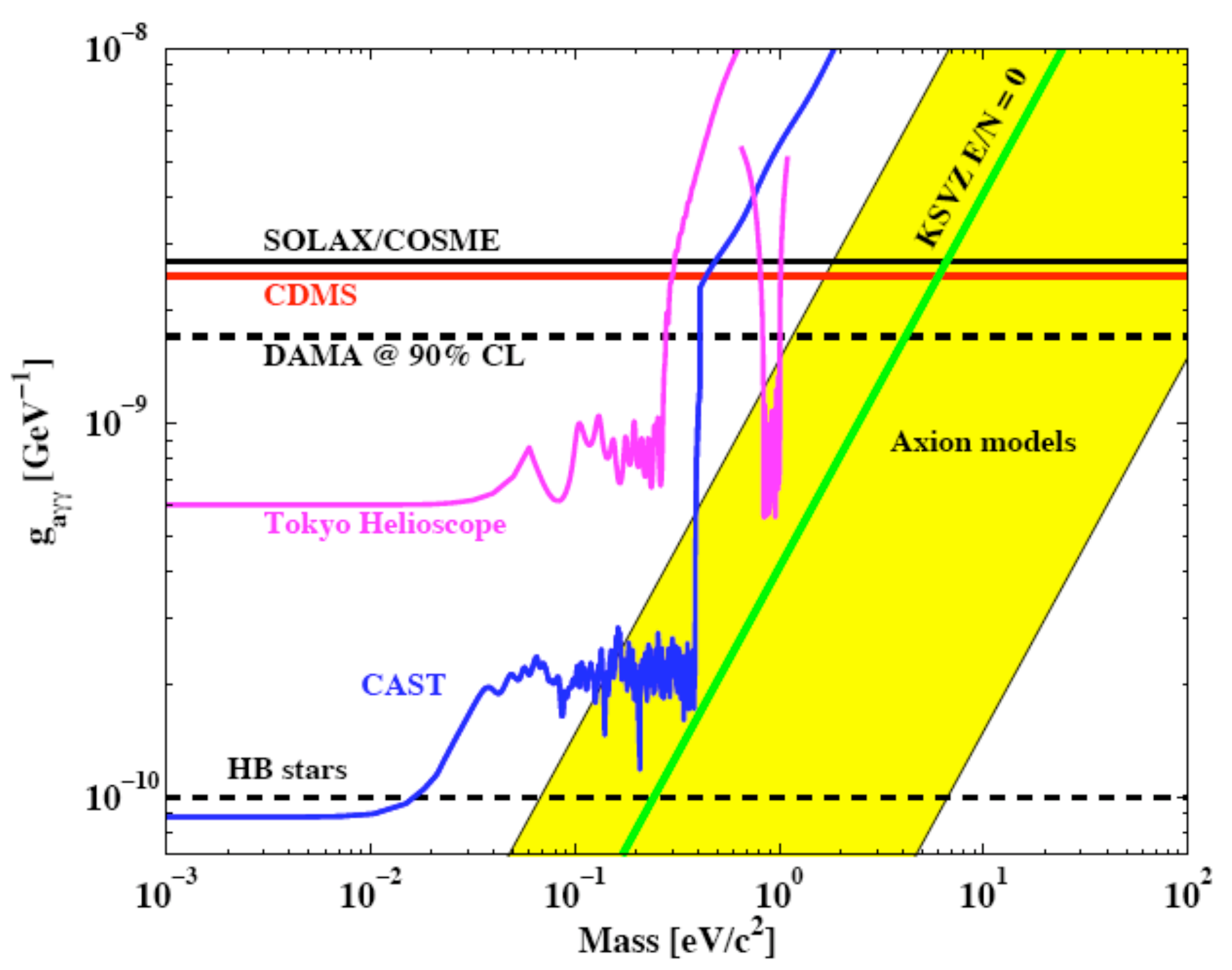}
\caption{The 95\% C.L. upper limit on the axion-photon coupling together with other experiment results.} 
\label{fig:solar_result}
\end{figure}

Each detector in a tower is rotated by $60^{0}$ with respect to the one above it. Each crystal's alignment, relative to true north, is known to $\pm3^{0}$. The expected calculated solar axion event rate in Ge for $g_{a\gamma\gamma}=10^{-8}$~GeV$^{-1}$ is shown in Fig.~\ref{fig:solar_rate}. An unbinned log-likelihood fit, similar to the described in section~\ref{sec:low_energy}, was used. The signal part of the fit included the expected event rate depicted in Fig.~\ref{fig:solar_rate} multiplied by the scale factor for the actual value of the axion-photon coupling. The scaling factor returned by the fit is $(1\pm1.5)\times10^{-3}$ and is consistent with zero~\cite{cdms_axions}. Thus, we did not observe any signatures of the solar axion conversion. The 95\% C.L. upper limit on the axion-photon coupling is shown in Fig.~\ref{fig:solar_result} together with other crystal search experiments (SOLAX/COSME and DAMA). 

\section{Summary}
The CDMS experiment has a world-leading upper limit on the WIMP-nucleon spin-independent cross section with a minimum of $4.6\times10^{-44}$ cm$^{2}$ at the 90\% CL for a 60 GeV/c$^{2}$ WIMP. It has the world's best sensitivity for WIMP masses above 44 GeV/c$^{2}$. Ongoing analysis of the remaining four runs with a raw exposure of $\sim$750 kg-days is in its final stage and will complete the 5-tower setup data analysis. Analysis of the low-energy electron-recoil spectrum sets a stringent experimental upper limit on the axio-electric coupling of $1.4\times10^{-12}$ at the 90\% CL for a 2.5~keV/c$^{2}$ axion. No excess in the counting rate above background in the 2--8.5 keV electron-recoil spectrum was found. In the solar axion search, the upper limit on the axion-photon coupling of $2.4\times10^{-9}$~GeV$^{-1}$ at the 95\% CL was set.

\bigskip 

\end{document}